\documentclass[prl,aps,twocolumn,floatfix]{revtex4}
\usepackage{epsfig}

\begin{document}
\title{Bistability between different dissipative solitons in nonlinear optics}

\author{U. Bortolozzo, L. Pastur, P.L. Ramazza}
\affiliation{Istituto Nazionale di Ottica Applicata, Largo E. Fermi 6, I50125 Florence, Italy}
\author{M. Tlidi}
\affiliation{Optique Nonlin\'eaire Th\`eorique, Universit\'e Libre de Bruxelles,
CP 231, Campus Plaine, 1050 Bruxelles, Belgium}
\author{G. Kozyreff}
\affiliation{Mathematical Institute, 24-29 St Giles', Oxford OX1 3LB, 
United Kingdom}

\date{\today}

\begin{abstract}

We report the observation of different localized structures 
coexisting for the same parameter values in an extended system. 
The experimental findings are carried out in a nonlinear optical interferometer, 
and are fully confirmed by numerical simulations. 
The existence of each kind of localized structure is put in relation to a 
corresponding delocalized pattern observed. 
Quantitative evaluation of the range of pump parameter allowing bistability 
between localized structures is given. The dependence of the phenomena on the 
other relevant parameters is discussed.

\vglue 0.3 truecm


\end{abstract}

\maketitle

Localization of structures is a widespread phenomenon occurring both in 
conservative and in dissipative systems. 
The study of this topic aims on the one hand to an understanding of the general 
conditions needed for the occurrence of localized structures (LS); and, on the 
other hand, to explore the potentialities offered by these objects in view of 
applications, such as information transmission and storage.
In this context, LS have been demonstrated in many diverse fields, including 
granular materials~\cite{Granular}, fluid dynamics~\cite{Fluid}, electroconvection 
in liquid crystals~\cite{Worms}, chemistry~\cite{Chemical} and 
nonlinear optics~\cite{Kivshar_Agarwal}.

With specific reference to optical systems, temporal solitons represent a 
classic and very active field of research.
Recently, spatial localized structures in extended nonlinear optical system
have been widely studied as well~\cite{Kivshar_Agarwal}. 

An important class of spatial LS, often referred to as cavity solitons, exists 
in many dissipative optical systems~\cite{Review_CS}. These solitons 
can be used as pixels for information storage or processing~\cite{TlidiMandel,FirthScroggie}.
Their existence has been experimentally demonstrated in several 
devices~\cite{LS_Anderson,LS_Darmstadt,LS_Noi,LS_Taranenko,LS_Nizza}, including 
miniaturized semiconductor resonators~\cite{LS_Taranenko,LS_Nizza}.

In this paper we present the experimental evidence of bistability between different 
localized structures in a dissipative nonlinear optical device. The two kind of 
solitons we find 
differ in shape, and are separated by a discrete gap in their peak intensity. If 
used as pixels for information storage, these LS represent three-state variables, 
instead of the common two-state variables ("bits") that a common soliton can 
encode. Consequently, use of these new solitons would lead to an increase of 
$log_2 3 \simeq 1.585$ for the information storable in a given area.

Localized structure bistability is observed in the presence of two modulational 
instabilities having different critical wavenumbers. The question of the interaction 
between these two instabilities  has been investigated in \cite{Tlidi} for 
semiconductor cavities.

The existence of bistable solitons has been predicted in the context of
modified nonlinear Schroedinger equations\cite{bistasol1,bistasol2}, describing
light propagation in conservative systems (e.g., optical fibers in the ideal,
lossless case). For the spatial dissipative solitons, multistability behaviour 
of either single or multi-peaked structures has been predicted numerically in
nonlinear optical
resonators~\cite{TlidiMandel,Michaelis_bistabili,Firth_bistabili}.
To our knowledge, an experimental observation of
bistability between different LS has not yet been given, neither in optics nor
in other fields.

Our experimental system is a nonlinear interferometer formed by a Liquid Crystal 
Light Valve (LCLV) with optical feedback. The phase $\varphi$ of an initially plane 
laser beam sent onto the device evolves as~\cite{Neubecker_Oppo}

\begin{equation}
\tau {\partial \varphi \over \partial t}=-(\varphi -\varphi _0) + l_d^2 \nabla^2
_{\perp} \varphi+ f(I_{fb}) 
\label{uno}
\end{equation}

\begin{equation}
I_{fb} = I_0 \mid e^{{il\nabla_{\perp}^2} / {2k_0}} (Be^{i\varphi} +1-B)
*sinc(\vec q_B \cdot \vec r){\mid}^2
\label{due}
\end{equation}

\begin{equation}
f(I_{fb}) = {\varphi}_{sat} (1-   e^{-{{\alpha I_{fb}}\over {\varphi}_{sat}} } )
\label{tre}
\end{equation}
where $l_d$ is the diffusion length of the LCLV, $I_{fb}$ the feedback intensity on 
the valve, and $\varphi_0$ the phase retardation induced by the valve on the input 
beam in the absence of feedback. 
In (2), $I_0$ is the input intensity, $B$ and $1-B$ are the fractions of the field 
along the principal directions of polarization of the cavity, 
$sinc(\vec q_B \cdot \vec r)$ represent the spatial filtering and $*$ denotes 
a convolution product. The parameters ${\varphi}_{sat}$, $\tau$ and $\alpha$ 
represent respectively the saturation phase value, the response time and the 
sensitivity of the LCLV. 
Finally, $l$ is the propagation distance inside the optical 
loop.

Eq. (3) describes the LCLV nonlinearity. In several circumstances, the
phenomena observed in this system can be understood by expanding the
exponential term in (3) to the first order, so that a Kerr approximation is
obtained~\cite{LS_Noi}. For the topics of interest here, it is instead
important to take into account the saturating character of the nonlinearity. 

It is well known that this system displays an extremely rich set of different 
dynamical behaviours~\cite{Akhmanov,Dalessandro,LS_Darmstadt,LS_Noi,Darmstadt_recente,
Papoff_recente}, and that extended as well as localized patterns can be 
formed in the transverse wavefront of the beam sent to the interferometer. 
For the present study, we set the system parameters at $l=30$ mm, $l_d \simeq
18 \  \mu m$, $B \simeq 0.52$, ${\varphi}_{sat} = 5$, $\varphi_0 = 0$ and the
spatial bandwidth $q_B=3.7$. Here and in the following, the spatial frequencies are 
expressed in units of the diffractive scale $q_{diffr}=2 \pi / \sqrt {2 \lambda l}$, 
where $\lambda$ is the optical wavelength.

\begin{figure}[!h]
\begin{center}
\leavevmode
\epsfxsize 80 mm
\epsfbox{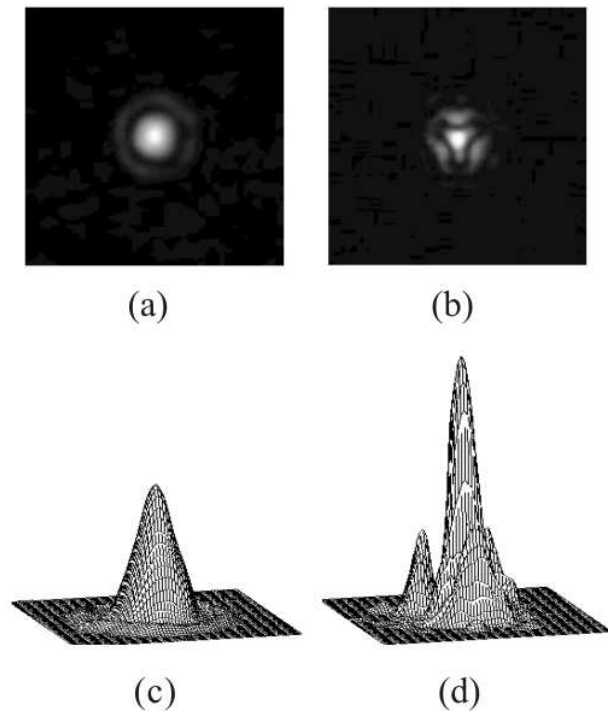}
\end{center}
\caption{(a), (c): Round soliton; (b), (d): Triangular soliton.}
\label{fig2}
\end{figure}

In our system, one type of LS has been shown to exist over very broad ranges of
parameters\cite{LS_Darmstadt,LS_Noi,CHAOS_Noi}. It has circular symmetry, with a bright central peak connected to
the dark background via a series of small amplitude oscillations along the
radial direction. Here, we report another completely different localized
solution exists for the same device. The two solitons, as observed in the
experiment, are shown in Fig.~\ref{fig2}. Numerical simulations faithfully 
reproduce the observations.

The most evident feature of the new LS is its triangular symmetry, 
observed both in the central peak and in the tails. Hence, in the following we
refer to these structures as Triangular Solitons (TS), and to the ones with
circular symmetry as round LS.

A second fundamental difference between triangular and round LS is their peak intensity. 
This can be appreciated in Fig. 1c, 1d. Finally the size of the central peak is 
substantially smaller for a TS than for a round LS, though the overall size of the two 
structures including the tails are comparable. 

The triangular soliton and the round LS shown in Fig. 1 are observed for
identical values of all the parameters,indicating bistability between the 
two structures.

The existence of solitons with triangular symmetry has been predicted recently 
reported in optical systems\cite{Michaelis_triangoli_e_Rosanov}.
However, no bistability between RS and TS is observed in these systems.

Each of these solitons can be switched on by an appropriate addressing pulse. Lower 
intensity pulses trigger a round LS, higher intensity ones a TS. In these 
regards, the observed weak sensitivity to the addressing intensity suggests the 
existence of large basins of attraction for each soliton; hence, 
the noise is not expected to be a serious limit in the use of these 
structures for information storage and processing tasks. 

We are now going to investigate the dependence of the phenomena observed on the 
pump intensity. The experimental state diagram of the system is shown 
in Fig. 2. 

\begin{figure}[!h]
\begin{center}
\leavevmode
\epsfxsize 65 mm
\epsfbox{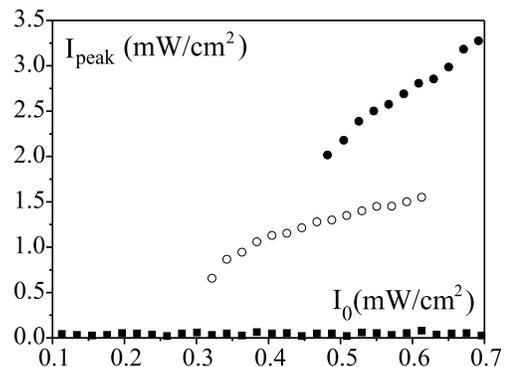}
\end{center}
\caption{Experimental state diagram of the system. Squares: low uniform state; empty 
circles: round localized structures; filled circles: triangular solitons.}
\label{fig4}
\end{figure}

Starting from a very low value of input intensity and gradually increasing it,
the lower uniform solution is the only state observed up to $I_0 \simeq 0.32$.
From this vaule on, round LS are observable, when addressed via proper initial
conditions.  At $I_0 \simeq 0.61$ $mW / cm^2$ the round LS loses its stability,
and the system jumps to the triangular solitons branch.  If, starting from a
TS, the pump is decreased, the structure remains stable down to $I_0 \simeq
0.47$ $mW / cm^2$, and then decays to a round LS.  If, instead, starting from
$I_0 \simeq 0.61$ $mW / cm^2$, the pump is increased, the TS exists up to  $I_0
\simeq 0.69$ $mW / cm^2$, and then destabilizes via a transition to a
delocalized irregular pattern. 

\begin{figure*}
\begin{center}
\leavevmode
\epsfxsize 130 mm
\epsfbox{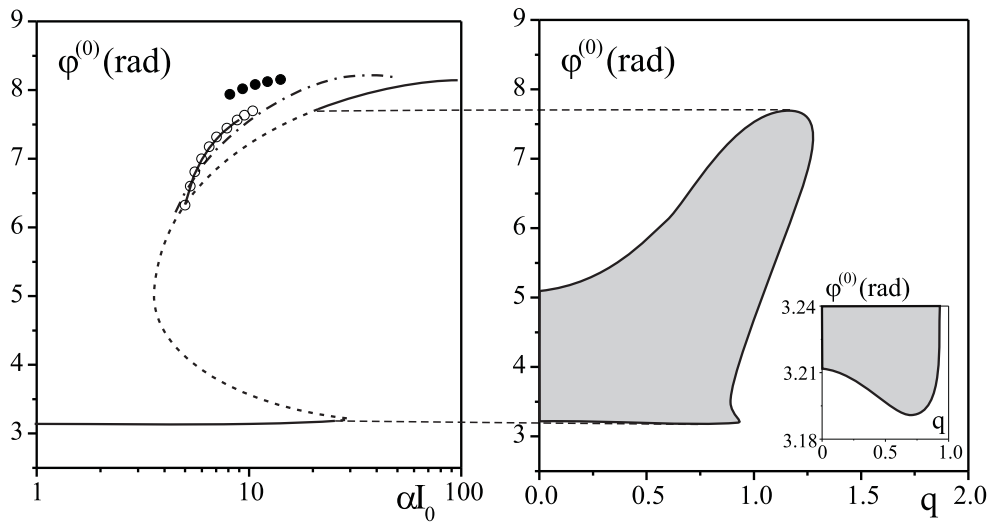}
\end{center}
\caption{Homogeneous steady state and its stability, together with localized and 
delocalized structure amplitude. For the values of parameters used, see text.
In (a): Continuous line: homogeneous solution, stable parts, and hexagons bifurcating 
at the lower critical point ($q=0.71$); dashed line: 
homogeneous solution, unstable parts; dash-dotted line: honeycombs bifurcating at the 
upper critical point ($q=1.15$); empty circles: 
round localized structures; filled circles: triangular solitons.}
\label{fig3}
\end{figure*}

The range of bistability between different LS
is therefore $0.47 \lesssim I_0 \lesssim 0.61$ $mW / cm^2$. Above $\simeq 2$
$mW / cm^2$ the system saturates to a uniform high output intensity state.

A useful insight about the origin of the LS observed can be gained from the observation of 
Fig. \ref{fig3}. Fig. 3(a) shows the homogeneous steady state (HSS) characteristic of 
the system $\varphi_0 \ vs \ \alpha I_0$, together with the amplitudes of the
localized and delocalized structures found in numerical simulations.  In Fig.
3(b) it is shown the instability region of the homogeneous steady state.

In the rest of the paper we mostly discuss numerical results; hence it is convenient 
to refer to the adimensional parameter $\alpha I_0$ as the pump, rather than 
to the intensity $I_0$ in physical units. In the conditions of the experiments here 
reported, $\alpha \simeq 18 \ cm^2 / mW$.

The lower branch of the HSS destabilizes at $\alpha I_0 \simeq 28$, for a
critical wavenumber $q_{c_1} \simeq 0.71$. The critical point is 
better visible in the inset of fig. 3b, showing an enlarged portion of the 
stability balloon. The bifurcation from this point is strongly
subcritical, and results in the formation of hexagons. The upper branch loses
stability at $\alpha I_0 \simeq 20$, $q_{c_2} \simeq 1.15$, with another
subcritical bifurcation leading to the appearence of honeycomb patterns.

Our simulations show that both the hexagons and the honeycombs are unstable 
within most of their existence region. Indeed, hexagons typically decay toward 
a collection of spatially well separated round LS; honeycombs are instead 
destabilized either towards a collection of triangular LS, or towards 
a situation of developed 
space-time chaos, depending on the value of $\alpha I_0$. In order to stabilize the 
periodic patterns and be able to track their amplitudes, we introduced a filter 
in Fourier space allowing only for the existence of the six fundamental modes 
needed to create hexagons and honeycombs, plus all their harmonic combination 
up to the cutoff frequency fixed at 3.7. In this way we obtain the amplitudes 
for the delocalized patterns shown in Fig. 3a. 

In the presence of the strongly subcritical patterns here observed, localized
structures can be formed in the parameter range where an homogeneous stable
steady state coexists, or is not far from, a patterned state~\cite{Riecke}.
These patterned states can be either stable or unstable for the
parameter values at which LS exist. In the latter case, LS can be interpreted
as residuals of some delocalized pattern that has lost its stability. 

In our case, the superposition of the round LS and of the hexagon amplitudes
seen in Fig. 3a clearly indicates the relation between the localized and
delocalized structure. As for the relation between triangular LS and
honeycombs, the results is less obvious from the data of Fig. 3a. We
conjecture however that such a connection exists also in this case. Our
conjecture is based both on the general argument which describes the LS as
local connections between a homogeneous and a patterned state; and, with
specific reference to our case, on the triangular symmetry which is common to
the localized structures and to the neighborhoods of local maxima in an
honeycomb lattice.

We remind that the role of an honeycomb pattern in giving rise to cavity solitons has been 
already pointed out in~\cite{Michaelis_bistabili}, for a model of semiconductor optical 
resonator. In that case, contrary to ours, dark solitons are formed by a dip connecting 
a stable bright state and a local minimum of the pattern.

We concentrated our quantitative experimental investigations on the dependence
of the phenomena observed on the pump intensity, while the other parameters
were kept fixed. Let us now discuss how the other parameters affect the
scenario. An exhaustive investigation of these dependencies is a demanding
task, which we do not face at the moment. We will here limit ourselves to
report about single parameter variations, with other parameters kept fixed at
the values used above.

The major effect of a decrease of the system bandwidth $q_B$ is a narrowing of
the stability range for both solitons; the effect is more marked for 
triangular solitons than for round LS. As a consequence, small values of $q_B$
lead to small ranges of soliton bistability. If $q_B$ is set to values higher
than $\simeq 4$, no substantial modifications occur with respect to the
situation described above.

The combined role of diffusion and diffraction is expressed by the adimensional 
parameter $\sigma \equiv (2 (l_d)^2 k_0 / l)^{1/2}$, giving the ratio of the diffusion 
to diffraction parameter. The quantity $\sigma$  can be viewed as an adimensional 
diffusion length. We are not in condition to have large excursion of $\sigma$ in the 
experiment. The numerics show that an increase of $\sigma$ with respect to the value 
used above leads to a shrinking of the stability range for both the round LS and 
the TS. The tendency is more marked for the triangular solitons, which have a
stronger content of high frequency components. 

The open loop phase $\varphi_0$ and the weights $B$ and 1-B of the feedback fields tune 
the S shape of the homogeneous steady state solution, and modify the location and 
critical wavenumbers of the modulational instabilities affecting the lower and upper 
branches. 
For the value of $B$ used above, bistability of LS is observed for 
$ - 0.4 \lesssim \varphi_0 \lesssim 0.1$. If $\varphi_0$ is kept fixed at 0, the 
two kinds of LS are bistable at $0.3 \lesssim B \lesssim 0.5$.

The saturation phase $\varphi_{sat}$ turns out also to be a very important 
parameter. We refer also in this case to numerical results, since the saturation parameter 
can be little varied in the experiment. If $\varphi_{sat}$ is lowered with respect to 
the value used above ($\varphi_{sat} = 5$), the stability range of TS moves to 
higher pump intensities, and these structures are no more observed when 
$\varphi_{sat} \lesssim 3.5$. If $\varphi_{sat}$ is increased, triangular solitons
continue to exist 
as well as round LS up to $\varphi_{sat} = 18$. The range of bistability has a 
broad maximum at $\varphi_{sat} \simeq 6$, and then shrinks, until disappearing 
at $\varphi_{sat} \simeq 18$. Above this value, the round LS still exist; in 
correspondence of the branch formerly corresponding to triangular solitons, now
only space-time chaotic delocalized patterns are observed. 

In summary, we have reported the evidence of bistability between different localized 
structures in a nonlinear optical interferometer. The observation of this 
phenomenon is completely new, and certainly more effort is required in order to obtain 
a full understanding of the sufficient and necessary conditions for its occurrence. 
Localized structure bistability occurs within a complex scenario of pattern forming 
events; still, it is clear that the phenomenon is robust with respect to parameter 
variations. This point, together with the generality of the mechanism lying at 
its basis, suggest that this phenomenon could be observed in systems of different 
nature. 

\vspace{5mm}

This work has been partially supported by the Interuniversity Pole Program of the 
Belgian Governmenti; by the Fonds National de la Recherce Scientifique; and by 
EU Contract HPRN-CT-2000-00158.


\begin{thebibliography}{100}

\bibitem{Granular}  P. Umbanhowar, F. Melo and H. Swinney, Nature {\bf 382}, 793 (1996).

\bibitem{Fluid}E. Moses, J. Fineberg and V. Steinberg, Phys. Rev. {\bf A35}, 2757 (1987); K. Lerman, E. Bodenschatz, D.S. Cannell, and G. Ahlers, 
Phys. Rev. Lett. {\bf70}, 3572 (1993).

\bibitem{Worms}  M. Dennin, G. Ahlers, and D.S. Cannell, Phys. Rev. Lett.
{\bf 77}, 2475 (1996)

\bibitem{Chemical} H. H. Rotermund, S. Jakubith, A. Von Oertzen and G. Ertl,
Phys. Rev. Lett. {\bf 66}, 3083 (1991); K.L. Lee, W.D. McCormick, Q. Ouyang
and H. Swinney, Science {\bf 261}, 192 (1993).

\bibitem{Kivshar_Agarwal} For a comprehensive review, see e.g. Y.S. Kivshar and
G.P. Agarwal, {\it Optical Solitons: from fibers to photonic crystals},
Academic Presss (2003).

\bibitem{Review_CS} See e.g. {\it Feature section on cavity solitons}, IEEE
Journal of Quantum Electronics 2003, {\bf 39}, n. 2 (2003).

\bibitem{TlidiMandel} M. Tlidi, P. Mandel and R. Lefever, Phys. Rev. Lett.
{\bf 73}, 640 (1994).

\bibitem{FirthScroggie} W.J. Firth and A.J. Scroggie., Phys. Rev. Lett. {\bf 76},
1623 (1996).

\bibitem{LS_Anderson} M. Saffman, D. Montgomery and D. Z. Anderson, Opt. Lett. {\bf 19}, 518 (1994).

\bibitem{LS_Darmstadt} A. Schreiber, B. Thuering, M. Kreuzer and T. Tschudi,
Opt. Comm. {\bf 136}, 415 (1997).

\bibitem{LS_Noi} P.L. Ramazza, S. Ducci, S. Boccaletti and F.T. Arecchi,
Journal of Optics B: Quantum and Semiclassical Optics {\bf 2}, 399 (2000).

\bibitem{LS_Taranenko} V.B. Taranenko, I. Ganne, R.J. Kuszelewicz and C.O.
Weiss, Phys. Rev. {\bf A 61} 063818 (2000).

\bibitem{LS_Nizza} S. Barland et al., Nature {\bf 419}, 699 (2002).

\bibitem{bistasol1} A.E. Kaplan, Phys. Rev. Lett. {\bf 55}, 1291 (1985); 
R.H. Enns, S.S. Rangnekar and A.E. Kaplan, Phys. Rev. {\bf A36}, 1270 (1987).

\bibitem{bistasol2} S. Gatz and J. Herrmann, Journal of Opt. Soc. Am. {\bf B8}, 
2296 (1991); W. Krolikowski and B. Luther-Davies, Opt. Lett. {\bf 17}, 1414 (1992).

\bibitem{Tlidi} G. Kozyreff, S.J. Chapman and M. Tlidi, Phys. Rev. {\bf E68}, 
015201 (2003).

\bibitem{Michaelis_bistabili} D. Michaelis, U. Peschel and F. Lederer, Phys.
Rev. {\bf A56}, R3366 (1997).

\bibitem{Firth_bistabili} J. McSloy, W.J. Firth, G.K. Harkness and G.L. Oppo,
Phys. Rev. {\bf E66}, 46606 (2002)

\bibitem{Neubecker_Oppo} R. Neubecker, G.L. Oppo, B. Thuering and T. Tschudi,
Phys. Rev.  {\bf A52}, 791 (1995).

\bibitem{Akhmanov} S. Akhmanov, M.A. Vorontsov and V.Y. Ivanov, JETP Lett. {\bf 47}, 707
(1988).

\bibitem{Dalessandro} G. D'Alessandro and W.J. Firth, Phys. Rev. Lett. {\bf
66}, 2597 (1991).

\bibitem{Darmstadt_recente} E. Benkler, M. Kreuzer, R. Neubecker and T. Tschudi,
Phys. Rev. Lett. {\bf 84}, 879 (2000).

\bibitem{Papoff_recente} S. Rankin, E. Yao, and F. Papoff Phys. Rev. {\bf A68},
013821 (2003)

\bibitem{Riecke} H. Riecke, "Localized structures in Pattern-Forming Systems" in
{\it Pattern Formation in Continuous and Coupled Systems}, ed. by
M. Golubitsky, D. Luss and S. Strogatz (IMA Volume 115, Springer, 1999),
p. 215.

\bibitem{CHAOS_Noi} Ramazza P.L., Boccaletti S., Bortolozzo U. and Arecchi
F.T., 2003 Chaos {\bf 13}, 335.

\bibitem{Michaelis_triangoli_e_Rosanov} D. Michaelis, U. Peschel, C. Etrich and
F. Lederer, IEEE journal of Quant. El. {\bf 39}, 255 (2003);S.V. Fedorov. N.N.
Rosanov, A.N. Shatsev, N.A. Veretenov and A.G. Vladimirov, {\it ibidem}.

\bibitem{Noi_Tuning} Ramazza P.L., Boccaletti S., Bortolozzo U., Ducci S., Benkler E. and
Arecchi F.T., 2002 Phys. Rev. {\bf E65}, 066204.

\end{thebibliography}
\end{document}